\documentclass[letterpaper]{article}       
\usepackage{authblk}
\usepackage{graphicx}
\graphicspath{{C:/Users/epaolini/Documents/Datos/ArchivosUNS/Trabajos/Guillermo/Experimental Astronomy 2011}}
\renewcommand\Affilfont{\small}
\begin{document}

\title{Achieving sub-electron readout noise \\
in Skipper CCDs}
\author[1,2,4]{Guillermo~Fern\'{a}ndez~Moroni}
\author[2]{Juan~Estrada}
\author[2]{Gustavo Cancelo}
\author[3]{Stephen E. Holland}
\author[1,5]{Eduardo E. Paolini}
\author[2]{H.~Thomas~Diehl}
\affil[1]{
Inst. de Inv. en Ing. El\'{e}ctrica (IIIE) ``Alfredo C. Desages''  \authorcr
Dto. Ing. El\'{e}ctrica  Comp., Universidad Nacional del Sur       \authorcr
Av. Alem 1253 - (B8000CPB) Bah\'{i}a Blanca, Argentina             \authorcr
\texttt{fmoroni.guillermo@gmail.com, epaolini@uns.edu.ar} \authorcr
\hspace{1cm}
}
\affil[2]{
Fermi National Accelerator Laboratory                       \authorcr
Kirk Road and Pine Street, Batavia, Il 60510, USA           \authorcr
\texttt{estrada@fnal.gov, cancelo@fnal.gov, diehl@fnal.gov} \authorcr
\hspace{1cm}
}
\affil[3]{
Lawrence Berkeley National Laboratory \authorcr  
1 Cyclotron Road, Berkeley, CA 94729, USA \authorcr
\texttt{seholland@lbl.gov} \authorcr
\hspace{1cm}
}
\affil[4]{
Comisi\'{o}n de Investigaciones Cient\'{i}ficas y T\'{e}cnicas (CONICET), Argentina \authorcr
\hspace{1cm}}
\affil[5]{
Comisi\'{o}n de Investigaciones Cient\'{i}ficas de la Prov. Buenos Aires (CIC), Argentina \authorcr
\hspace{1cm}
}
\date{}
\maketitle

\begin{abstract}
The readout noise for Charge-Coupled Devices (CCDs) has been the main limitation when using these detectors for measuring small amplitude signals. A readout system for a new scientific, low noise CCD is presented in this paper. The Skipper CCD architecture, and its advantages for low noise applications are discussed. A technique for obtaining sub-electron readout noise levels is presented, and its noise and signal characteristics are derived. We demonstrate a very low readout noise of $0.2e^{-}$ RMS. Also, we show the results using the detector in a low-energy X-ray detection experiment. \authorcr
\\
\hspace{1cm}
\end{abstract}

\newpage
\section{Introduction}
\label{intro}
Since its invention as a memory device, CCDs have been widely used as
imaging detectors because of their capability to obtain high resolution
digital images of objects placed in its line of sight. Astronomy and
spectroscopy applications have been using this technology since its invention \cite{Boyle2010,Smith2010} encouraging continuous improvements \cite{Flaugher2006,McLean2008}. Nowadays, scientific CCDs are developed with high performance
characteristics for visible and infrared light detection, with quantum
efficiencies above 90\%, wide dynamic range with capability to handle
signals producing more than 100,000 $e^{-}$ RMS, and pixel size around
10 $\mu $m \cite{Janesick2001}. However the readout noise (RN) caused by the
CCD output stage is still the main limitation when these detectors are used
for low signal detection or energy discrimination of slightly different signals.

The RN is an error in the pixel value caused by electrical noise added to
the CCD video signal. The main source of this electronic noise is the CCD
output amplifier but the readout system and the bias and clocks signals used
for charge collection and transfer do also contribute 
\cite{McLean2008,Janesick2001}. The readout system is in charge of recovering and
digitizing the pixel value encoded as the difference between two constant
levels in the video signal, therefore electronic noise fluctuations altering
those levels are interpreted as variations in the collected charge.

In this paper we describe a new Skipper CCD with sub-electron RN. The Skipper technology was developed in the  '40  \cite{Chandler1990,Janesick1990} and uses floating gate output stage that allows for
multiple sampling of the same pixel \cite{Wen1974}. Some promising
experimental results are presented, achieving RN levels of $0.2~e^{-}$ RMS.

The paper is organized as follows. In Section \ref{Sec:Skipper}, a brief
description of the Skipper CCD is provided, and its output stage is compared
with the floating diffusion output stage commonly used for standard CCDs.
The capability of non-destructive, multiple reading of the same pixel is
also described. The typical sources of video signal noise are described in
Section \ref{Sec:OutputNoiseAndReadout}, and the readout systems of standard
and Skipper CCDs are reviewed. Their frequency response is derived, and the
Skipper noise reduction technique is analyzed. Some experimental results,
including applications for X-ray detection are described in Section \ref%
{Sec:Experimental}. Finally, some concluding remarks are provided in Section %
\ref{Sec:Conclu}.

\section{Skipper CCD detector}
\label{Sec:Skipper}

\begin{figure}
\centering
\resizebox{0.6\hsize}{!}{\includegraphics*{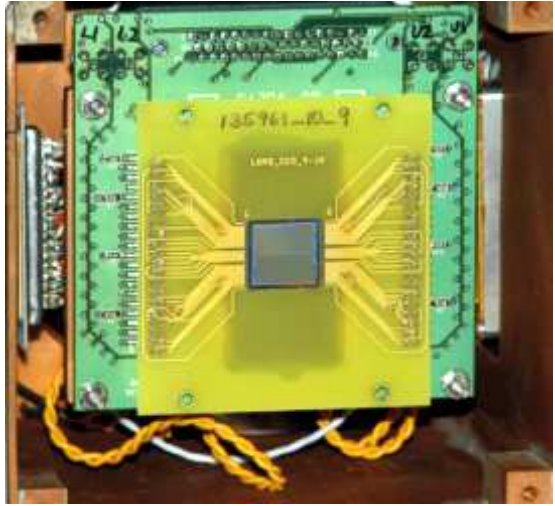}}
\caption{Skipper CCD in a testing package mounted in a dewar at Fermilab.}
\label{fig:Skipper}
\end{figure}%

\begin{figure}
\centering
\includegraphics[scale=0.8]{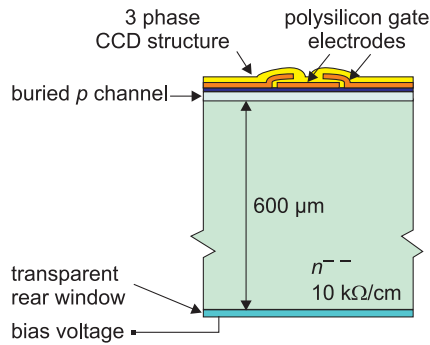}
\caption{Cross section of a pixel in the Skipper CCD (from \protect{\cite{Holland2003}}).}
\label{fig:SkipperPixel}
\end{figure}

This detector (Fig.\thinspace \ref{fig:Skipper}) was developed by Stephen
Holland in Berkeley Laboratory in 2010. It is a \emph{p}%
-channel front illuminated CCD fabricated on high-resistivity, \emph{n}-type
silicon. A low density impurities implant in the \emph{n}-region together
with a reverse bias voltage in a ohmic contact in the wafer back side allows
to operate the detector with a large fully depleted region increasing the
quantum efficiency for infrared light. The resistivity is about 10000 ohm/cm
and 40V is a common substrate voltage for the detector. Using this
combination, it is possible to obtain a depletion region width around 250 $%
\mu $m in comparison with the 20 $\mu $m of standard detectors \cite%
{Holland2003}. A cross section of the CCD for one pixel is shown in
Fig.\thinspace \ref{fig:SkipperPixel}.

\begin{figure}
\centering
\includegraphics{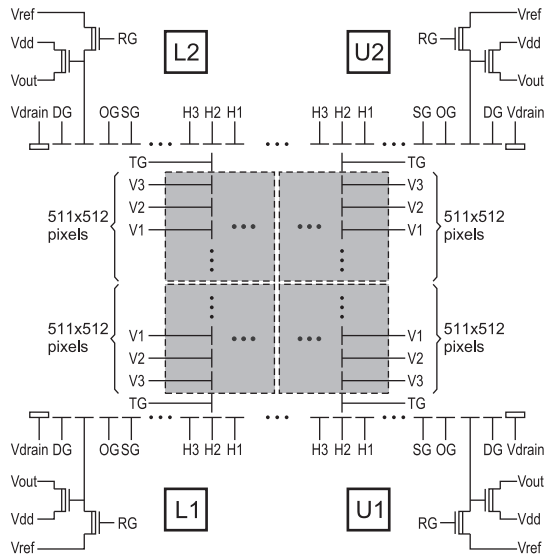}
\caption{Skipper CCD Architecture. Each of the four amplifiers (U2, U1, L1 and L2) handles a 512 by 512 pixel region.}
\label{fig:SkipperCircuit}
\end{figure}

The detector is composed by an array of 1022 by 1024 pixels with a pixel
width of 15 $\mu $m. The sensors are grouped in four regions (L1, L2, U1 and
U2) which are read by four amplifiers placed in each corner of the chip. The
horizontal registers that move the charge to the output stage are placed in
two sides of the detector (Fig.\thinspace \ref{fig:SkipperCircuit}). Both
horizontal and vertical gates are arranged in a three phase configuration 
\cite{Janesick2001} where V1, V2 and V3 move the charge vertically, and H1,
H2 and H3 transfer it horizontally to the output amplifier. In the output
stage the charge is moved by the summing well gate (SG), output gate (OG) and and dump gate (DG) clocks (see Section \ref{Sec:SkipperOperation}).

\begin{figure}
\centering
\includegraphics[scale=0.8]{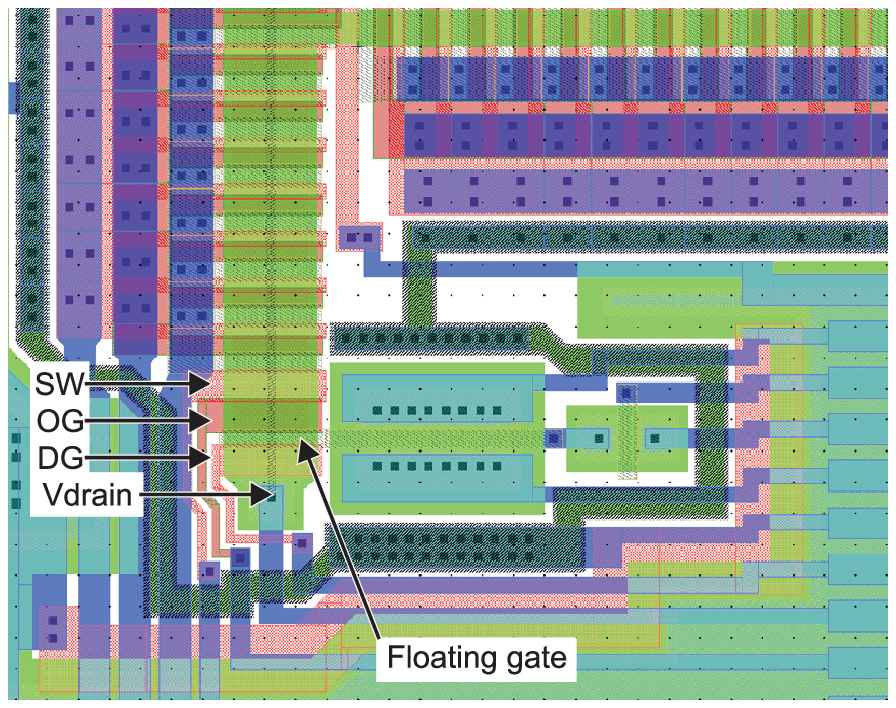}
\caption{L2 amplifier layout. The summing well gate (SG), output gate (OG), drain gate (DG), drain contact (Vdrain) and floating gate are shown.}
\label{fig:SkipperLayout}
\end{figure}

The charge packet is transferred to the sense node, where the output
amplifier in a common drain configuration drives the video signal to the
readout system. A distinctive feature of the Skipper CCD is its floating
gate output stage \cite{Chandler1990,Wen1974} that allows for non
destructive charge measurements, instead of the common floating diffusion
output stage used in standards CCDs. The layout of the output stage of the
L2 amplifier is shown in Fig.\thinspace \ref{fig:SkipperLayout}.

In the following section the sensitivity of the traditional floating
diffusion output stage is compared to the floating gate of the Skipper CCD.

\subsection{Floating diffusion output stage}

This is the typical output configuration in standards CCDs since it provides
the highest sensitivity (voltage increment at the CCD output per hole
collected) because the output amplifier gate is connected directly to the
CCD \emph{p}-channel providing a minimum capacitance. When a package of
charge is transferred to the floating diffusion node, it produces a voltage $%
V=Q/C$, where $Q$ is the charge packet transferred and $C$ is the total
capacitance between the node and ground given by%
\begin{equation}
C=C_{J}+C_{P}+C_{MOS},
\end{equation}%
where $C_{J}$ is the channel-bulk junction capacitance for the sense node
region, $C_{P}$ is the parasitic capacitance between the diffusion and
surrounding CCD gates, and $C_{MOS}$ is the output transistor input
capacitance, which depends on its dimensions and polarization. The signal-to-noise ratio at the output increases with low $C$ values, so the effect of
the noise in the pixel value is strongly related to $C$ \cite{Janesick2001}.

\subsection{Floating gate output stage}

In this configuration the output transistor gate is connected to the channel
through a gate with a silicon oxide insulating layer that provides a stop
barrier for holes when they are measured, so the charge remains in the sense
node after it is measured \cite{Wen1974}. With properly clock timing, the
charge can be moved back and forth between the SG region and the sense node,
allowing for multiple samples of the same pixel (see next section). The
sensitivity becomes a critical aspect of the design because of the extra
capacitance added by the floating gate that acts as a capacitive divisor
between the sense node and the output transistor gate \cite{Janesick2001}.
In this case the voltage signal at the transistor gate is%
\begin{equation}
V=\frac{Q}{C_{J}+C_{P}+C_{MOS}+(C_{P}+C_{J})C_{MOS}/C_{OX}},
\end{equation}%
where $C_{OX}$ is the new oxide capacitance of the floating gate. The new
term $(C_{P}+C_{J})C_{MOS}/C_{OX}$ in the equivalent capacitance reduces the
amplitude of the voltage signal available at the output node. From this
equation it seems that increasing $C_{OX}$ is a possible solution for
reducing the extra capacitance, but it does not take into account the
increase of capacitance $C_{J}$ as the gate area is increased. In actual
designs, $C_{OX}$ is larger than $C_{MOS}$, and smaller gate areas are used
to reduce the junction capacitance for a net increase of the sensitivity.
However, small gate area reduces the full well capacity (the maximum number
of holes handled by the device) so the size of floating gate area results in
a trade off between sensitivity and full well capacity.

To explore how this variables are related and its impact on the sensor
performance, the Skipper CCD has been designed with different gate areas for
each amplifier (L1: 216 $\mu $m$^{2}$, L2: 96 $\mu $m$^{2}$, U1: 60 $\mu $m$%
^{2}$, U2: 96 $\mu $m$^{2}$).

Despite the extra capacitance, the sensitivity observed for our Skipper
detector is similar to detectors using floating diffusion configuration,
under the same testing conditions. The charge-to-voltage conversion factor
was estimated at $3.5$ ${\mu}$V$/e^{-}$ \cite{Holland2003}.

\subsection{Skipper operation}
\label{Sec:SkipperOperation}

\begin{figure}
\centering
\includegraphics[scale=0.9]{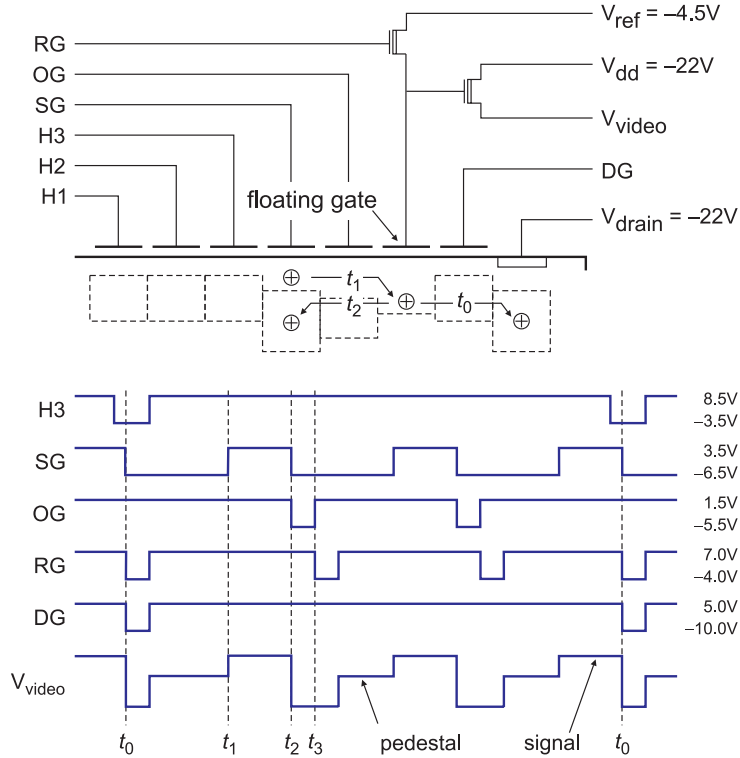}
\caption{Skipper output stage and timing.}
\label{fig:SkipperTiming}
\end{figure}%

The main advantage of the floating gate output stage of the Skipper CCD is
its capability of performing multiple pixel readings with minimum effect on
the storage charge. The flow of the charge is controlled by the summing gate
clock (SG), output gate clock (OG), and dump gate clock (DG) and the timing
of these signals for reading out the same pixel three times are shown in
Fig.\thinspace \ref{fig:SkipperTiming}. In $t_{0}$ the charge package is
placed under the last H3 gate at the end of the serial register. The SG
clock goes down attracting the charge pocket under its gate. After a few
microseconds H3 goes high and the charge remains completely stored under SG.
At the same time DG and the reset gate (RG) signals go down to remove the
previous pixel charge from the channel and to set the new reference voltage
for the current pixel. In $t_{1}$ the charge is transferred from the SG to
the sense node through the OG. The SG is set high increasing its channel
potential above the OG barrier so the charge moves to the potential well set
by the last reset pulse in the sense node. At this point the charge produces
a change in the reference level at the output transistor gate that is
reflected as a step increment in the output video signal. The skipper cycle
continues when the holes are transferred backward again to the SG in $t_{2}$%
. This process is achieved moving OG and SG to low level. As the SG sets a
lower channel potential all the charge goes directly under its surface and
the OG acts like a potential barrier to confine the charge, in this case,
under the SG. In $t_{3}$, after the charge is moved out from the sense node,
the RG resets the floating gate to the $V_{ref}$ voltage resulting in a new
reference level for the new pixel sample. Using this sequence the pixel
value is encoded as the difference of two constant levels. The reference
level is called the \emph{pedestal level}, and the second one the \emph{%
signal level}.

By repeating $t_{1}$, $t_{2}$ and $t_{3}$ cyclically, several measurements
of the same pixel can be performed. After reading the desired number of
samples, the pixel reading procedure is completed when the charge is removed
from the channel and the floating gate is reset by the DG and RG pulse in $%
t_{0}$ (of the next pixel), respectively. When DG goes down, it leaves a
potential path between the sense node and the $V_{drain}$ ohmic contact for
channel discharging.

The key point for the noise reduction algorithm is that the noise is less
correlated among samples, and therefore simple averaging can be used for
noise reduction.

\section{Output noise and readout systems}
\label{Sec:OutputNoiseAndReadout}

Regardless of the type of CCD output stage, the pixel measurement is
differentially encoded between two constant levels in the video signal, and
the amount of charge can be calculated simply by subtracting them. However,
as the signal is corrupted by electronic noise originated mainly in the
output transistor, more sophisticated readout methods should be used to
achieve low RN. There are two different mechanisms that produce signal
fluctuation, and each one is characterized by a distinctive power spectrum
density (PSD) and a different RN contribution in the pixel value \cite%
{Janesick2001,Chandler1990}. These noise sources are:

\begin{itemize}
\item white (Johnson) noise, caused by random fluctuation of free charges
that are part of the transistor current;

\item $1/f$ (flicker) noise, caused by traps in the output transistor
channel that catch and release moving charge for relatively long periods,
producing a noise with more power at low frequencies.
\end{itemize}

Figure \ref{fig:SkipperCCDRFN} shows the noise PSD (red line) for the CCD
video signal. Both noises mechanisms are clearly recognizable: high power
components at low frequencies for the $1/f$ noise, and a flat PSD at high
frequency as part of the white noise.

The main task of the readout system is to recover the pixel value while
rejecting the noise. In the following subsections the readout system for
standard and Skipper CCDs are briefly reviewed, and their noise and signal
performance are derived.

\begin{figure}
\centering
\includegraphics{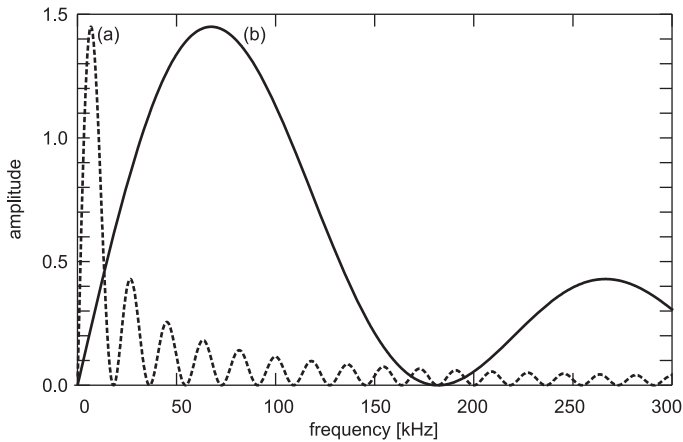}
\caption{Frequency response of the DSI readout system for $T = 110$ $\mu$s (a) and $T = 11$ $\mu$s (b).}
\label{fig:RF_DSI}
\end{figure}%

\subsection{Standard CCD readout system}

The dual slope integrator (DSI) readout system \cite{wey1990} has been
widely used because it provides an optimal filtering for white noise \cite%
{McLean2008}. Instead of taking just one sample of each constant level, the
DSI performs an integration during a given time interval of both the
pedestal and the signal levels. The pixel charge is obtained as the
difference between the results of the integration in both regions.

Using this description it is possible to derive the frequency response of
the DSI to analyze the effects of noise in the image pixel. As the pixel
value is the difference between the results of the pedestal and signal level
integrations, the DSI impulse response is%
\begin{equation}
h_{DSI}(t)=\left\{ 
\begin{array}{rl}
A/(T_T/2), \quad &  ~~~~~~~~~~t_{0}\leq t\leq t_{0}+T_T/2, \\ 
-A/(T_T/2),\quad &  t_{0}+T_T/2\leq t\leq t_{0}+T_T, \\ 
0, \quad&   \mbox{otherwise,}%
\end{array}%
\right.
\end{equation}%
where $A$ is an arbitrary integration gain, $t_{0}$ is an arbitrary time
instant when the pedestal begins, and $T_T$ is the total integration time, reparted equally between the signal and pedestal levels. This
model assumes that transients in the video signal from the CCD that occur in
the transition between pedestal and signal intervals, and viceversa, and
others caused by clock leakages, etc. have been removed. 
From now on, we consider $A=1$ and $t_{0}=0$ for simplicity. Then the module of the frequency response is given by%
\begin{equation}
|H_{DSI}(f)|=\frac{4 A}{\pi T_T f}\sin ^{2}\left(\frac{\pi T_T}{2} f \right).
\end{equation}

Plots of $|H_{DSI}(f)|$ for two different integration times are shown in
Fig.\thinspace \ref{fig:RF_DSI}. The frequency response for $%
T_T=110~\mu $s is plotted as a black dashed curve (a), and for $T_T=11~\mu $s as a
black solid trace (b). When the total integration time $T_T$ is increased, the frequency response curve shifts to a lower frequency band, and its bandwidth decreases, (each lobe gets thinner) reducing the effective bandwidth of the readout system. Therefore, the incidence of white noise on the pixel value is reduced. However, this RN reduction is limited by the $1/f$ noise, because of the shift of the frequency response curve to lower frequencies where the noise PSD is higher. 

There is an optimal total integration
time $T_T$ that minimizes the contribution of both white noise and $1/f$ noise.
 The minimum noise level achieved by increasing the integration time in
detectors with floating diffusion output stage and DSI readout system is
around $2e^{-}$ RMS \cite{Janesick2001,Holland2003}.

\subsection{Skipper CCD readout system}
\label{Sec:SkipperReadout}

The multiple readout technique allows to go below the $2e^{-}$ RMS limit given
by the $1/f$ noise and get sub-electron RN. After the multiple sampling, the
readout system averages the samples for each pixel to get its final value.
As each pixel is the average of $N$ samples and each sample is read using
the DSI technique, the impulse response $h_{skp}(t)$ of the readout system is the linear
combination of time shifted copies of $h_s(t)$:%
\begin{equation}
	h_{skp}(t)=\frac{1}{N}\sum\limits_{n=0}^{N-1}h_s\left( t-n(\tau +T_T/N)\right) ,
\end{equation}%
where
\begin{equation}
h_s(t)=\left\{ 
\begin{array}{rl}
A(2N/T_T), \quad &  ~~~~~~~~~~~~~~~t_{0}+\tau \leq t\leq t_{0}+\tau +T_T/(2N), \\ 
-A(2N/T_T),\quad &  t_{0}+\tau +T_T/(2N)\leq t\leq t_{0}+\tau +T_T/N, \\ 
0, \quad&   \mbox{otherwise,}%
\end{array}%
\right.
\end{equation}%
and $\tau$ is the time interval between samples. In this case each pixel is read $N$ times during an interval $T_T/N$, but the total
integration time employed by the readout system for reading one pixel is again $T_T$. The $2N$ jumps in the impulse response, alternating between positive and negative levels, causes extra rejection for correlated low frequency components compared to the standard DSI.

Using Fourier transform properties, and assuming $t_0=\tau=0$ for simplicity, the module of the frequency response for the skipper readout system results in%
\begin{equation}
|H_{skp}(f)|=\frac{4A}{\pi T_T f}\sin ^{2}\left(\frac{\pi T_T}{2N} f \right)\left\vert \frac{\sin (\pi T_T f)}{\sin(\pi T_T f/N)}\right\vert .
\end{equation}

\begin{figure}
\centering
\includegraphics{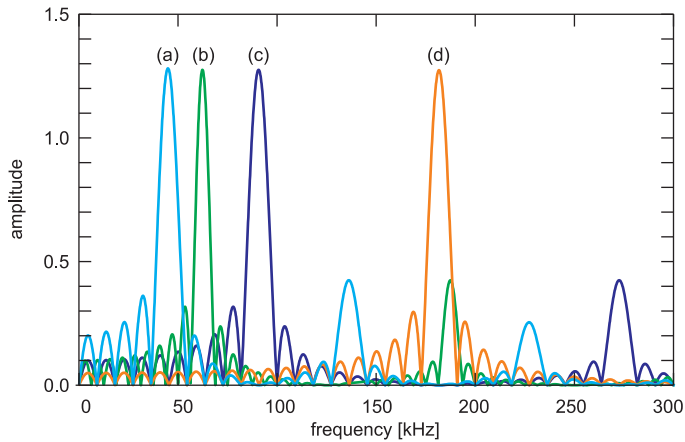}
\caption{Frequency response of the Skipper readout system for $T_T = 110$ $\mu$s and $N=5$ (a), $N=10$ (c) and $N=20$ (d), and for $T_T=160$ $\mu$s, and $N=10$ (b).}
\label{fig:Skipper_RF}
\end{figure}

The performance of the Skipper CCD readout system can be explained with the help of Fig.\thinspace \ref{fig:Skipper_RF}. Curves (a)-(d) depict the frequency response of the readout system for different values of $T_T$ and $N$. Curves (a), (c) and (d) correspond to a constant integration time $T_T$, but increasing number of averaged samples $N$ (5, 10, 20, respectivley). As the total integration time $T_T$ remains constant, the bandwidth of each response is approximately the same because it is dominated by the zeros of the $\sin(\pi T_T f)$. Increasing the number of averages shifts the frequency response moves to higher frequencies, and the gain for low frequencies is reduced. 

The curve (b) corresponds to a  total integration time $T_T=160$ $\mu$s and $N=10$ averages. As the total integration time $T_T$ is increased, the bandwidth is narrower than in the case of $T=110$  $\mu$s and $N=10$ shown in curve (c); the larger integration time $T_T$ displaces the main lobe to a lower frequency, and the lower number of averages $N$ increases the low-frequency gain.

These results show that to achieve maximum performance of the Skipper readout system, 1) a total integration time $T_T$ as large as possible must be used to narrow the bandwidth of the system, and 2) the number $N$ of measurements must be chosen to obtain maximum rejection at low frequency and dispace the main lobe of the frequency response to higher frequencies avoiding the $1/f$ limitation. However, it should be clear that the extra RN reduction is paid by an extra readout time, so this should be consider for each aplication.

Although in
theory there is no limitation in RN reduction by increasing the number $N$ of
averaged samples, there is a practical limit imposed by some losses in the
charge transfer. However, in our experiments we have not found significant
degradation of the signal even when sampling each pixel 1600 times.

Therefore, choosing apropriate values for the total integration time $T_T$ and the number of averaged samples $N$, the Skipper CCD\ readout system can be adjusted to reject the $1/f$ noise and provide an optimal filtering of white noise because the integration performed over the signal. 

\subsection{Comparison of DSI and Skipper performance}
\begin{figure}
\centering
\includegraphics{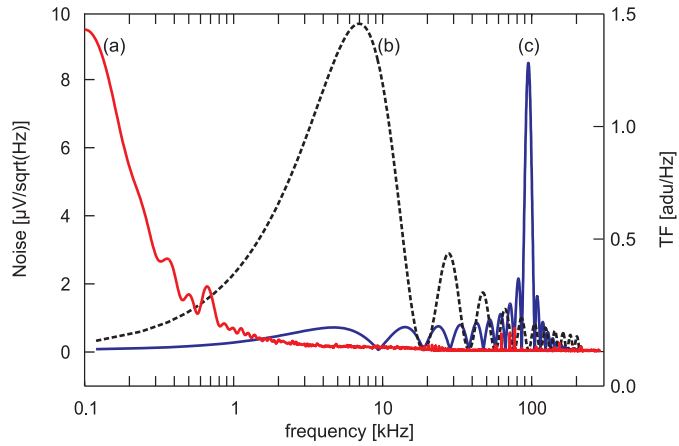}
\caption{Left axis: CCD noise PSD (a). Right axis: frequency response of DSI readout system for $T_T=110$ $\mu$s (b) and of the Skipper readout system for $T_T = 110$ $\mu$s and $N=10$ (c).}
\label{fig:SkipperCCDRFN}
\end{figure}

The main advantage of the Skipper readout system is perceived
when its frequency response is compared to the standard DSI frequency
response for the same time integration interval, i.e., using the same total amount
of time for reading each pixel. In Fig.\thinspace \ref{fig:SkipperCCDRFN} the frequency response  $|H_{DSI}(f)|$ of the DSI readout system (curve (b)) is plotted  for $T_T=110$ $\mu$s, together with the frequency response of Skipper readout system for $T_T=110$ $\mu $s, and $N=10$ (curve (c)). These responses are the same as curve (a) in Fig.\thinspace \ref{fig:RF_DSI}, and curve (c) in Fig.\thinspace \ref{fig:Skipper_RF}, respectively, but plotted using a logarithmic frequency axis. Also in Fig.\thinspace \ref{fig:SkipperCCDRFN} the PSD of CCD noise is depicted as curve (a).
This figure reveals that the Skipper CCD readout system has a higher rejection for high power, low frequency components allowing a RN reduction even in presence of $1/f$ noise, and this rejection can be increased by merely augmenting the number $N$ of averaged samples. 

In other words, low frequency noise
rejection is achieved and the $1/f$ limitation is removed. 

\section{Experimental results}
\label{Sec:Experimental}

In this section, the behavior of the noise-reduction technique for the
skipper readout system to achieve sub-electron RN levels is analyzed. This
extremely low background noise is not only of interest \emph{per se}, but
also of paramount importance in certain applications: we show how this
technology can be used to reduce uncertainty in low energy X-ray detection
in a controlled X-ray detection experiment.

The readout system utilized in this experiment was originally developed for
the DECAM project (Dark Energy Camera, \cite{honscheid2008}) at Fermilab \cite{Diehl2008,Estrada2006} and it was modified to handle the
skipper technology. The detector was tested at one of the Fermilab's
laboratories. The CCD was installed in an aluminum dewar and it was run at $%
143$ ${{}^\circ}$K to avoid dark current generation. 
Each sample was read with an integration time of $T=10.4$ $\mu $s and then the total integration time for each pixel is $T_T=2NT$. This integration time does not take into account the duration of the signal to pedestal and pedestal to signal pulses. 
Therefore, the actual reading time may be a small percentage larger than $T_T$.

\begin{figure}
\centering
\includegraphics[scale=1.05]{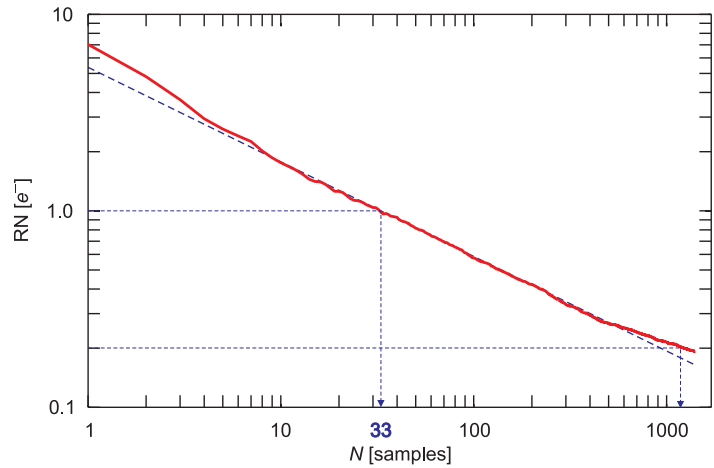}
\caption{Skipper CCD RN as a function of the number of averaged samples $N$. Continuous line: RN measured from images. Dashed line: theoretical white noise fit for the image RN.}
\label{fig:SkipperCCDRNs}
\end{figure}%

\subsection{Background noise measurements}
As stated in Section \ref{Sec:SkipperReadout}, the RN level can be reduced
by increasing the number $N$ of samples taken from each pixel. A series of
measurements in the deterctor's overscan region were performed to
measure the background noise of the CCD. This test is useful to determine
the number of samples $N$ required for obtaining sub-electron RN levels.
Figure \ref{fig:SkipperCCDRNs} shows the RMS noise in electrons ($e^{-}$ RMS%
) as a function of the number $N$ of samples averaged for each pixel. The
continuous line depicts the RN noise measured from the images, and the
dashed line is the theoretical white noise reduction fitted for those
points. The differences are caused by $1/f$ and other correlated noise
components present in the video signal. The figure shows that as the number $%
N$ of averaged samples increases, the measured RN decreases, following the
white noise tendency. This effect corroborates the analysis of Section \ref%
{Sec:SkipperReadout}, revealing that augmenting the number of averaged
samples of each pixels increases the rejection of low frequency ($1/f$)
noise, and as a result the main noise contribution is given by the white
noise component.

\begin{figure}
\centering
\includegraphics{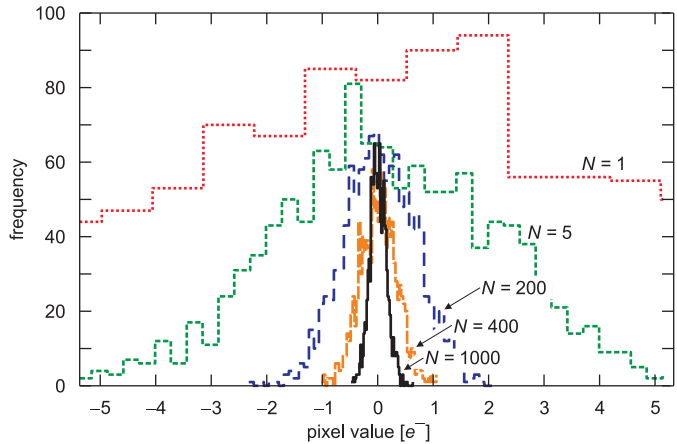}
\caption{Background pixel histograms from Skipper CCD images for different number of averaged samples $N$ per pixel.}
\label{fig:SkipperCCDRN}
\end{figure}%

The figure reveals that sub-electron RN can obtained for $N=33$ and a RN = $%
0.2e^{-}$ RMS is achieved when averaging $N=1227$ samples. The same results are
depicted in a more intuitive form using histograms in Fig.\thinspace \ref%
{fig:SkipperCCDRN}. The width of the Gaussian distribution is greatly
reduced by the averaging of samples, and for $N=200$ all the RN values are
practically contained within the $2e^{-}$ RMS segment allowing for very low
signal detection. The RN noise level achieved for different number of averaged samples is summarized in Table \ref{tab:RNnoise}.

\begin{table}
\begin{center}
\caption{RN noise vs. number of averaged samples $N$}
\label{tab:RNnoise}       
\begin{tabular}{ll}
\hline\noalign{\smallskip}
$N$ & RN [$e^{-}$ RMS]  \\
\noalign{\smallskip}\hline\noalign{\smallskip}
1    & 7.00 \\
5    & 2.60 \\
200  & 0.40 \\
400  & 0.29 \\
1400 & 0.19 \\
\noalign{\smallskip}\hline
\end{tabular}
\end{center}
\end{table}

\subsection{X-ray experiment}

To verify the skipper functionality for low RN applications, it was applied to reduce uncertainty in low-energy X-ray detection in a controlled X-ray detection experiment. Jointly with the detector a Fe55 X-ray source and a teflon target were installed in the dewar. The detector is mounted in front of the target, and this in front of the source. A shielding is placed between the source and the CCD for stopping direct X-rays.

\begin{figure}
\centering
\includegraphics{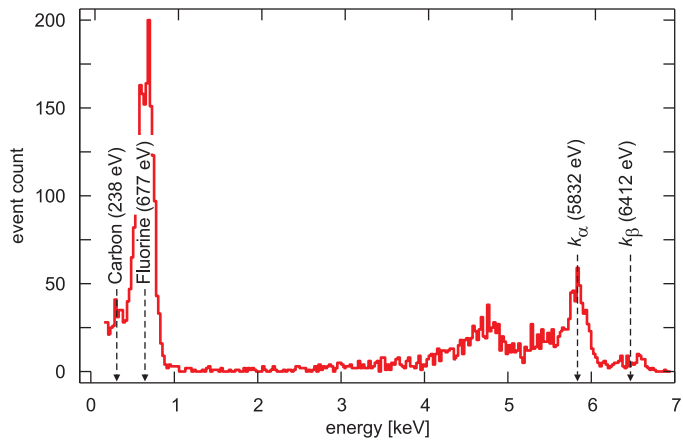}
\caption{Histogram of the X-ray energy detected by the Skipper CCD.}
\label{fig:Xray}
\end{figure}

The Fe55 X-ray source produces two different energy rays from Mn: 5832~eV ($K_\alpha$) and 6412~eV ($K_\beta$). Both X-rays hit the  teflon with carbon and fluorine, which  emits  lower energy X-ray by fluorescence. Each atom has a precise energy pattern of emitting photons \cite{Janesick1990}. For the carbon and fluorine the most probable emitted X-rays are at energies of  277~eV and 677~eV, respectively. These low-energy X-rays are detected by the CCD together with some high energy X-rays coming directly from the source that get to cross the shielding. Figure \ref{fig:Xray} shows the histogram of the X-rays detected by the Skipper CCD. The largest  peak is given by the fluorine that is the most common element in the Teflon composition, so it has the most probable interaction. Both spurious peaks of the source are present to a lesser extent at high energies. The small peak at the left of the fluorine peak is the carbon peak, which is pretty small because of  the low energy X-rays have very small penetration in the material, so most of them get trapped before leaving the target in the same teflon material (or in the chip layers on the top of the silicon for front-illuminated CCDs). Due to its low noise background, the Skipper CCD multiple-sample technique allows for a much cleaner images and improves the low-energy hit detection. 

\section{Conclusions}
\label{Sec:Conclu}

A readout technique for sub-electron noise level
measurements was devised for a new type of CCD imager suitable for low
background noise applications. This method is based on the capability of the
Skipper CCD of admitting several non-destructive measurements of the charge
storage in each pixel. The noise and signal characteristics of the proposed
readout system are derived, and compared to traditional methods used for
typical CCDs (DSI).

Experimental results confirm the expected behavior of the system, and its
application to reduce uncertainty in low energy X-ray detection in a
controlled X-ray detection experiment reveals the utility of this technique.

It is expected in a near future to update the detector system of the direct
dark matter search using CCDs (DAMIC) experiment with this technology \cite%
{Estrada2008}.

\end{document}